%% file: spin-4PM-dissipative.tex
\documentclass[twocolumn,superscriptaddress,amsmath, amssymb, amsfonts,preprintnumbers,aps,prd,longbibliography,nofootinbib]{revtex4-1}

\renewcommand{\sec}[1]{\textit{#1. --- }}

\usepackage{graphicx}
\usepackage{dcolumn}
\usepackage{bm}
 \usepackage{amstext}
 \usepackage{amssymb}
\usepackage{subcaption}
 \usepackage{amsmath}
 \usepackage{graphicx}
 \usepackage{color}
 \usepackage{bbold}
 \usepackage{delimset} 
\usepackage[colorlinks=true]{hyperref}
\usepackage{orcidlink}

\usetikzlibrary{decorations.pathmorphing}
\usetikzlibrary{decorations.markings}
\usetikzlibrary{positioning, shapes, snakes, arrows}
 \usepackage{tikz-feynman}

\tikzset{ 
	graviton/.style={line width=.8pt, -latex,decorate, decoration={snake, segment length=4pt,amplitude=1.8pt, pre length=.1cm, post length=.25cm}},
	worldline/.style={gray, line width=1pt},
	worldlineBold/.style={black, line width=.6pt},
        background/.style={black,dotted,line width=1pt},
	zUndirected/.style={line width=1pt},
	zParticle/.style={line width=1pt,postaction={decorate},decoration={markings,mark=at position .6 with {\arrow[#1]{latex}}}},
	zParticleF/.style={line width=1pt,postaction={decorate}},
	cscalar/.style={line width=1pt,postaction={decorate},decoration={markings,mark=at position .6 with {\arrow[#1]{latex}}}},
	cscalar2/.style={line width=1pt,postaction={decorate},decoration={markings,mark=at position .8 with {\arrow[#1]{latex}}}},
	photon/.style={line width =.8pt, decorate, decoration={snake, segment length=3pt, amplitude=1.8pt,  pre length=.1cm, post length=.1cm}},
	 mid arrow/.style={postaction={decorate,decoration={
        markings,
        mark=at position .5 with {\arrow[#1]{latex}}}}} ,
        worlddot/.style={dotted, line width=.8pt},
	worlddot2/.style={dotted, line width=1pt}   }

\DeclareFontFamily{OT1}{pzc}{}
\DeclareFontShape{OT1}{pzc}{m}{it}{<-> s * [1.350] pzcmi7t}{}
\DeclareMathAlphabet{\mathpzc}{OT1}{pzc}{m}{it}

\def\cN{\mathcal{N}}
\def\cO{\mathcal{O}}

\def\eps{\epsilon}

\def\d{\mathrm{d}}
\def\D{\mathrm{D}}

\def\dd{\delta}

\def\d{\mathrm{d}}
\def\eps{\epsilon}

\def\nn{\nonumber}

\def\eqn#1{eq.~\eqref{#1}}

\def\Fig#1{Fig.~{\ref{#1}}}

\def\Rcite#1{Ref.~\cite{#1}}
\def\Rcites#1{Refs.~\cite{#1}}

\newcommand*\Bell{\ensuremath{\boldsymbol\ell}}

\newcommand{\vev}[1]{\langle #1\rangle}

\newcommand{\be}{\begin{equation}}
\newcommand{\ee}{\end{equation}}
\newcommand{\ba}{\begin{align}}
\newcommand{\ea}{\end{align}}
\ifx\genfrac\sdflkaj\else\fi
\newcommand{\sfrac}[2]{{\textstyle\frac{#1}{#2}}}

\newcommand{\mO}{\mathcal{O}}
\newcommand{\mn}{{\mu\nu}}

\newcommand*{\vct}[1]{\boldsymbol{#1}}

\begin{document}

\preprint{HU-EP-23/47-RTG}

\title{Dissipative scattering of spinning black holes at fourth post-Minkowskian order
}

\author{Gustav Uhre Jakobsen\,\orcidlink{0000-0001-9743-0442}} 
\affiliation{%
Institut f\"ur Physik und IRIS Adlershof, Humboldt Universit\"at zu Berlin,
Zum Gro{\ss}en Windkanal 2, 12489 Berlin, Germany
}
 \affiliation{Max Planck Institut f\"ur Gravitationsphysik (Albert Einstein Institut), Am M\"uhlenberg 1, 14476 Potsdam, Germany}

\author{Gustav Mogull\,\orcidlink{0000-0003-3070-5717}}
\affiliation{%
Institut f\"ur Physik und IRIS Adlershof, Humboldt Universit\"at zu Berlin,
Zum Gro{\ss}en Windkanal 2, 12489 Berlin, Germany
}
 \affiliation{Max Planck Institut f\"ur Gravitationsphysik (Albert Einstein Institut), Am M\"uhlenberg 1, 14476 Potsdam, Germany}
 
 \author{Jan Plefka\,\orcidlink{0000-0003-2883-7825}} 
\affiliation{%
Institut f\"ur Physik und IRIS Adlershof, Humboldt Universit\"at zu Berlin,
Zum Gro{\ss}en Windkanal 2, 12489 Berlin, Germany
}

\author{Benjamin Sauer\,\orcidlink{0000-0002-2071-257X}} 
\affiliation{%
Institut f\"ur Physik und IRIS Adlershof, Humboldt Universit\"at zu Berlin,
Zum Gro{\ss}en Windkanal 2, 12489 Berlin, Germany
}

\begin{abstract}
We compute the radiation reacted momentum impulse $\Delta p_i^\mu$, spin kick $\Delta S_i^\mu$,
and scattering angle $\theta$ between two scattered spinning massive bodies
(black holes or neutron stars) using the
$\cN=1$ supersymmetric worldline quantum field theory formalism
up to fourth post-Minkowskian (4PM) order.
Our calculation confirms the state-of-the-art non-spinning results,
and extends them to include spin-orbit effects.
Advanced multi-loop Feynman integral technology including differential equations and the
method of regions are applied and extended to deal with the retarded propagators arising in a causal description of the scattering dynamics.
From these results we determine a complete set of radiative fluxes at sub-leading PM order:
the 4PM radiated four-momentum and, via linear response,
the 3PM radiated angular momentum, both again including spin-orbit effects.
\end{abstract}
 
\maketitle 

Predicting the gravitational waveforms emitted during the encounter of two spinning
black holes (BHs) or neutron stars (NSs) \cite{LIGOScientific:2016aoc, *LIGOScientific:2017vwq,*LIGOScientific:2021djp}
at highest precision is a central challenge in todays
gravitational physics. Reaching the necessary high precision,
which stretches far beyond the present state-of-the-art in analytical and numerical gravitational wave physics,
is mandatory \cite{Purrer:2019jcp} for the scientific reach of
the third generation of detectors \cite{LISA:2017pwj,*Punturo:2010zz,*Ballmer:2022uxx} that are scheduled to go online in the 2030s \cite{Maggiore:2019uih,*Barausse:2020rsu}.
To achieve this, a combination of perturbative and numerical approaches --- post-Newtonian \cite{Blanchet:2013haa,*Porto:2016pyg,*Levi:2018nxp}, 
post-Minkowskian (PM) \cite{Kosower:2022yvp,*Bjerrum-Bohr:2022blt,*Buonanno:2022pgc,*DiVecchia:2023frv,*Jakobsen:2023oow}, self-force \cite{Mino:1996nk,*Poisson:2011nh,*Barack:2018yvs,*Gralla:2021qaf},  effective-one-body resummations \cite{Buonanno:1998gg,*Buonanno:2000ef} and
numerical relativity \cite{Pretorius:2005gq,*Boyle:2019kee,*Damour:2014afa} --- are currently being used to solve the relativistic two-body problem.
By importing techniques from perturbative quantum field theory (QFT),
great progress has been made in PM
wherein one assumes weak gravitational fields but arbitrarily fast-moving bodies.

By analogy to collider physics, the PM scenario considers the \emph{scattering} 
of BHs or NSs \cite{Kovacs:1978eu,*Westpfahl:1979gu,*Bel:1981be,*Damour:2017zjx,*Hopper:2022rwo}. On the one hand, this is well-motivated for mergers with
highly eccentric orbits; on the other hand the scattering data may
be used to inform models of the bound-state problem \cite{Cheung:2018wkq,*Kalin:2019rwq,*Kalin:2019inp,*Saketh:2021sri,*Gonzo:2023goe,Cho:2021arx}. As long as
the objects' separation is large compared to their intrinsic sizes, 
the BHs or NSs have an effective description in terms of a massive point particle
coupled to Einstein's theory of gravity 
\cite{Goldberger:2004jt}.
Based on this effective worldline approach two-body scattering observables (deflections and
Bremsstrahlung waveforms)
have recently been computed at high orders in the PM expansion,
including spin and tidal effects
\cite{Kalin:2020mvi,*Kalin:2020fhe,*Kalin:2020lmz,*Dlapa:2021npj,*Dlapa:2021vgp,*Liu:2021zxr,*Mougiakakos:2021ckm,*Riva:2021vnj,*Mougiakakos:2022sic,*Riva:2022fru,*Mogull:2020sak,*Jakobsen:2021smu,*Jakobsen:2021lvp,Jakobsen:2021zvh,Jakobsen:2022fcj,Jakobsen:2022psy,Shi:2021qsb,*Bastianelli:2021nbs,*Comberiati:2022cpm,*Wang:2022ntx}.
In the non-spinning case,
the present state-of-the-art for the impulse and scattering angle is
4PM ($G^{4}$) order including radiation-reaction effects \cite{Dlapa:2022lmu,Dlapa:2023hsl}.
Complementary QFT-based approaches,
based on scattering amplitudes~\cite{Neill:2013wsa,*Luna:2017dtq,*Kosower:2018adc,*Cristofoli:2021vyo,*Bjerrum-Bohr:2013bxa,*Bjerrum-Bohr:2018xdl,*Bern:2019nnu,*Bern:2019crd,*Bjerrum-Bohr:2021wwt,*Cheung:2020gyp,*Bjerrum-Bohr:2021din,*DiVecchia:2020ymx,*DiVecchia:2021bdo,*DiVecchia:2021ndb,*DiVecchia:2022piu,*Heissenberg:2022tsn,*Damour:2020tta,*Herrmann:2021tct,*Damgaard:2019lfh,*Damgaard:2019lfh,*Damgaard:2021ipf,*Damgaard:2023vnx,*Aoude:2020onz,*AccettulliHuber:2020dal,*Brandhuber:2021eyq},
have similarly achieved a 4PM precision for the impulse in the non-spinning case~\cite{Damgaard:2023ttc},
plus next-to-leading order results for the scattering waveform
\cite{Brandhuber:2023hhy,*Herderschee:2023fxh,*Georgoudis:2023lgf,*Elkhidir:2023dco,*Caron-Huot:2023vxl}.
In the spinning case results exist for the impulse at 3PM order up to quadratic spins~\cite{FebresCordero:2022jts}
and at 2PM order for higher spins~\cite{Bern:2022kto,*Aoude:2022thd,*Aoude:2023vdk}.
Other approaches to PM spin effects are \cite{Vines:2017hyw,*Bini:2017xzy,*Bini:2018ywr,*Guevara:2017csg,*Vines:2018gqi,*Guevara:2018wpp,*Chung:2018kqs,*Guevara:2019fsj,*Chung:2019duq,*Guevara:2020xjx}.

In fact the spins of the compact objects (next to the masses)
are prime observables in the observed binary black hole and neutron star mergers
\cite{Ramos-Buades:2019uvh,*Chiaramello:2020ehz,*Nagar:2021gss,*Liu:2021pkr,*Khalil:2021txt,*Hinderer:2017jcs,*Islam:2021mha}. 
Fascinatingly, the most efficient way to capture the spins of BHs or NSs in the worldline approach 
is by upgrading to a superparticle worldline theory, an approach inspired by superstring
theory \cite{Jakobsen:2021zvh}.
The generalization to spin-spin interactions uses $\mathcal{N}=2$ supersymmetry.
Use of this worldline quantum field theory (WQFT) approach has 
given rise to state-of-the-art scattering observables, including spin-orbit and spin-spin effects
at 3PM order~\cite{Jakobsen:2022fcj,Jakobsen:2022zsx} and most recently also
spin-orbit at 4PM order for the conservative sector of BH or NS scattering \cite{Jakobsen:2023ndj}.

In this Letter we report on the upgrade of these 4PM results to the full dissipative, 
radiation-reacted observables.
To compute them, we use the  Schwinger-Keldysh (in-in) initial value formulation of the WQFT that induces
the use of retarded propagators and a causality flow in the diagrammatic expansion \cite{Jakobsen:2022psy}.
Using these results, we are then able to provide a complete set of radiative fluxes
at sub-leading order in the PM expansion ---
$G^4$ for the radiated four-momentum $P_{\rm rad}^\mu$
and $G^3$ for the radiated angular momentum $J_{\rm rad}^\mu$.
The latter is inferred from the 4PM scattering observables using linear response
\cite{Bini:2012ji,Bini:2021gat}.
Finally, we demonstrate the 4PM total scattering angle's
natural decomposition into terms coming from different regions of integration.
All of our results are included in an ancillary file attached to the \texttt{arXiv}
submission of this Letter.

\sec{Scattering of spinning bodies}
The scenario we are interested in is one of two initially well-separated spinning massive bodies $i=1,2$ scattering off each other.
In the far past,  the gravitational field is weak and the bodies move on straight line
trajectories $x_{i}^{\mu}(\tau)=b_{i}^{\mu} + v_{i}^{\mu}\tau$;
they carry four-momenta $p_i^\mu=m_iv_i^\mu$ (with masses $m_i^2=p_i^2$ and boost factor $\gamma=v_1\cdot v_2$)
and intrinsic angular momenta (spins) $S_i^{\mu\nu}$.
The total initial angular momentum of the two-body system is then given by
\begin{align}
  \begin{aligned}
    J^{\mu\nu}&=L^{\mu\nu}+S_1^{\mu\nu}+S_2^{\mu\nu}\,,\\
    L^{\mu\nu}&=2b_1^{[\mu}p_1^{\nu]}+2b_2^{[\mu}p_2^{\nu]}\,.
  \end{aligned}
\end{align}
However, these angular momentum tensors depend on the choice of center of each (extended) body and the coordinate origin.
To resolve this, we find it convenient to introduce the spin vectors
\begin{subequations}
  \begin{align}
    L^\mu&=\sfrac12{\eps^\mu}_{\nu\rho\sigma}L^{\nu\rho}\hat P^\sigma=
    -\sfrac{1}{E} {\eps^\mu}_{\nu\rho\sigma} b^\nu p_1^\rho p_2^\sigma\,,\label{eq:Lvec}\\
    S_i^\mu&=m_ia_i^\mu=\sfrac12{\eps^\mu}_{\nu\rho\sigma}S_i^{\nu\rho}v_i^\sigma\,,\label{eq:aVec}
  \end{align}
\end{subequations}
where the impact parameter $b^\mu=|b|\hat b^\mu$ is the part of $b_2^\mu-b_1^\mu$ orthogonal to both initial momenta $p_i^\mu$
and $P^\mu=E\hat{P}^\mu=p_1^\mu+p_2^\mu$ is the total initial momentum. ``Hatted'' vectors are unit-normalized.
In addition we introduce the total angular momentum vector in the center-of-mass (CoM) frame
\begin{align}\label{eq:jVec}
  \begin{aligned}
    J^\mu=&\sfrac12{\eps^\mu}_{\nu\rho\sigma}J^{\nu\rho} \hat P^\sigma
    =
    \,L^\mu
    +
    2
    \hat P_\nu
    \sum_i v_i^{[\nu} S_i^{\mu]}\,  .
  \end{aligned}
\end{align}
We also define the symmetric mass ratio $\nu=\mu/M=m_1m_2/M^2$, the total mass $M=m_1+m_2$,
the total energy $E=M\Gamma=M\sqrt{1+2\nu(\gamma-1)}$ and $\delta=(m_2-m_1)/M$.
The 4PM observables that we compute are the change in momentum (impulse) $\Delta p_i^\mu$,
and of spin (spin kick) $\Delta S_i^\mu$, including dissipative losses.

\sec{Supersymmetric in-in WQFT formalism}
The effective $\mathcal{N}=1$ supersymmetric worldline theory of Kerr BHs or spinning NSs with masses $m_i$, trajectories $x_{i}^{\mu}(\tau)$ and anti-commuting vectors $\psi_i^\mu(\tau)$ carrying the spin degrees of freedom takes the compact form~\cite{Jakobsen:2021zvh} 
\begin{align}\label{eq:action}
S=-\sum_{i=1}^{2}m_{i}\!\int\!\d\tau\bigg[\sfrac{1}{2}g_{\mu\nu}\dot x_i^{\mu}\dot x_i^{\nu}
  \!+\! i\psi_{i,\mu}\!\frac{\D\psi_i^\mu}{\d\tau}\!\bigg] + S_{\rm EH}
\end{align}
in proper time gauge $\dot x_{i}^{2}=1$ (we use a mostly minus metric) and with covariant derivative $\D\psi_i^\mu/\d\tau$.
The bulk Einstein-Hilbert action $S_{\rm EH}$ includes a de Donder gauge-fixing term and we employ dimensional regularization with $D=4-2\epsilon$.
The fields are expanded around their initial-state asymptotic motion:
\begin{align}
  \begin{aligned}\label{backgroundexp}
    x_i^\mu(\tau) &= b_i^\mu \!+\! v_i^\mu \tau \!+\! z_i^\mu(\tau)\,, &
    \psi^\mu_i(\tau) &= \Psi^\mu_i\!+\!{\psi'}_i^\mu(\tau)\,,
  \end{aligned}
\end{align}
with perturbative deflections $z_i^\mu(\tau)$ and ${\psi}_i^{\prime\mu}(\tau)$
and the initial $\Psi^{a}_i$ related to the spin tensors as $S_i^{\mu\nu}=-i m_i \Psi_i^{\mu}\Psi_i^{\nu}$.
Similarly, working a post-Minkowskian (PM) expansion we expand the metric $g_\mn=\eta_\mn +\sqrt{32\pi G}h_\mn$.
Details on the supersymmetric worldline formalism may be found in \Rcites{Jakobsen:2021zvh,Jakobsen:2023ndj}.

Our objective is now to find perturbative-in-$G$ solutions to the equations of motion for the 
superfield deflections $Z_i^\mu(\tau)=\{z_i^\mu(\tau),\psi_i^{\prime\mu}(\tau)\}$.
This can efficiently be accomplished using the WQFT formalism,
which generates the perturbations in a diagrammatic fashion.
The fields $Z_i^\mu(\tau)$ and $h_{\mu\nu}(x)$ are promoted to propagating degrees of freedom, with propagators
\begin{subequations}
  \begin{align}
    \begin{tikzpicture}[baseline={(current bounding box.center)}]
      \coordinate (in) at (-0.6,0);
      \coordinate (out) at (1.4,0);
      \coordinate (x) at (-.2,0);
      \coordinate (y) at (1.0,0);
      \draw [zUndirected] (x) -- (y) node [midway, below] {$\omega,n$} node [midway, above] {$\rightarrow$};
      \draw [background] (in) -- (x);
      \draw [background] (y) -- (out);
      \draw [fill] (x) circle (.08) node [above] {$\mu$};
      \draw [fill] (y) circle (.08) node [above] {$\nu$};
    \end{tikzpicture}&=\frac{-i\eta^{\mu\nu}}{m_i(\omega+i0^+)^{n}}\,
    \begin{cases}
   \text{$n=2$ for $z_{i}^{\mu}\,,$}\\  \text{$n=1$ for $\psi_{i}^{\prime\mu}\,,$}
    \end{cases} 
    \\
      \label{eq:gravProp}
      \begin{tikzpicture}[baseline={(current bounding box.center)}]
    \begin{feynman}
    \coordinate (x) at (-.7,0);
    \coordinate (y) at (0.5,0);
    \draw [photon] (x) -- (y) node [midway, below] {$k$} node [midway, above] {$\rightarrow$};
    \draw [fill] (x) circle (.08) node [above] {$\mu\nu$};
    \draw [fill] (y) circle (.08) node [above] {$\rho\sigma$};
    \end{feynman}
    \end{tikzpicture}
          &=\frac{i(\eta_{\mu(\rho}\eta_{\sigma)\nu}-\sfrac1{D-2}\eta_{\mu\nu}\eta_{\rho\sigma})
          }{k^{2}+ \text{sgn}{(k^{0})}i 0^{+}}\, .
  \end{align}
  \end{subequations}
Arrows on the propagators indicate causality flow:
we use retarded propagators in order to fix boundary conditions at past infinity.
From a QFT perspective, this is formally accomplished  using the Schwinger-Keldysh in-in formalism~\cite{Jakobsen:2022psy,Kalin:2022hph}.
Lower-multiplicity worldline vertex rules have been exposed explicitly in \cite{Jakobsen:2021zvh} where a generic vertex
couples $n$ gravitons to $m$ worldline fields and conserves the energy on the worldline, see \cite{Jakobsen:2023ndj}.
The WQFT formalism exploits the fact that \emph{tree-level one-point functions}
$\vev{Z_{i}^{\mu}(\tau)}$ solve the \emph{classical} equations of motion \cite{Boulware:1968zz} ---
trivializing the classical limit.

\sec{WQFT integrand construction}
Automated construction of the 4PM integrands is done recursively using Berends-Giele type relations \cite{Jakobsen:2023ndj}. Inserting the
Feynman rules into the generated trees is done with {\tt FORM} \cite{Ruijl:2017dtg}.
We face 529 with spin graphs contributing to the 4PM impulse and 253 contributing to the 4PM spin kick.
The 4PM observables fall into two classes defined by their mass dependence: 
test-body contributions with linear mass dependence, $m_{1} m_{2}^{4}$ or   $m_{1}^{4} m_{2}$, and 
comparable-mass contributions, $m_{1}^{2} m_{2}^{3}$ or   $m_{1}^{3} m_{2}^{2}$.

The integrand initially consists of a large sum of tensor integrals with integrations on both worldline energies and graviton momenta.
Conservation of energy, however, results in unconstrained integration only on the space-like components of the momenta.
At 4PM order this results in three-loop integrals, depending on the  
momentum transfer $q$ (the Fourier transform of the impact parameter) and
the parameter $\gamma=v_{1}\cdot v_{2}$. After scaling out $|q|$
we are faced with a collection of three-loop single parameter tensor integrals 
which we reduce to scalar integrals using Veltman-Passarino reduction~\cite{Jakobsen:2023ndj}.

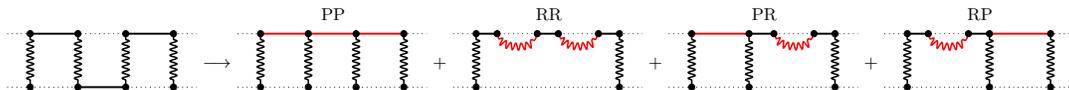
\begin{figure*}[ht!]
  \input{Graphs_Emergence_RRPP}
  \caption{\small
    Example of a 4PM master integral decomposing into the four regions in the static limit.
    There is no distinction between the velocities any longer, which is why the middle
     worldline propagator effectively moves up, taking the gravitons along or not. This
     results in active propagators (marked in red) in the (PP), (RR), (PR) regions as marked.
   }
  \label{fig:GraphsEmergence}
\end{figure*}

\begin{figure*}[ht!]
  \input{RRP_Graph}
  \caption{\small
    The boundary integral topologies:
    The first seven graphs result in twelve boundary master integrals (MIs) and belong to the (PP) region. In the (RR) region the double mushroom graph gives two MIs and the last two single mushroom graphs result in six MIs in the (PR) region. 
   }
  \label{fig:rrGraphs}
\end{figure*}
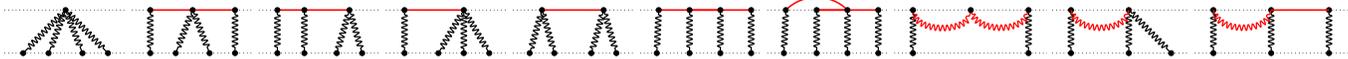

\sec{Reduction to master integrals}
The complete tensor-reduced spin-orbit integrands for the impulse and spin kick
consist of some $10^{5}$ integrals.
In the comparable-mass sector the main family is given by
\begin{subequations}\label{eq:jFamily}
  \begin{align}
    J^{(\sigma_1,\sigma_2,\sigma_3,\sigma_4,\sigma_5)}_{n_1,n_2,\ldots,n_{12}}
    :=
    \int_{\ell_1,\ell_2,\ell_3}\!\!\!\!
    \frac{
      \dd(\ell_1\!\cdot\! v_{1})
      \dd(\ell_2\!\cdot\! v_{1})
      \dd(\ell_3\!\cdot\! v_{2})
    }{
      D_1^{n_1}
      D_2^{n_2}\cdots
      D_{12}^{n_{12}}
    }
  \end{align}
  with ($j=1,2,3$, $k=1,2$)
  \begin{align}
    &
    D_k
    =
    \ell_{k}\cdot v_{2} +\sigma_{k}i0^+
    \,,\,\,
    D_3
    =
    \ell_{3}\cdot v_{1} +\sigma_{3}i0^+\,,
    \nn\\
    &
    D_{3+k}
    =
    (\ell_k-\ell_3)^2+\sigma_{3+k}\,\text{sgn}\,(\ell_k^0-\ell_3^0)\,i0^+
    \,,\,\,\\
    &D_{6}
    =
    (\ell_1-\ell_2)^2
    \, ,
   D_{6+j}
   =\ell_j^2
   \,,\,\,
   D_{9+j}
   =(\ell_j+q)^2
   \,,\nn
  \end{align}
\end{subequations}
and $q\cdot v_i=0$.
Additionally we need for a few integrals in this sector the second family $I^{[ 2 ]}$;
the definition of this as well as the definition of the family for the test-body integrals $I^{[ 1  ]}$. They are  given in the supplementary material
\eqn{eq:iFamily}  \cite{Jakobsen:2023ndj}.
We split each family up into even ($b$-type) or odd ($v$-type)
in the number of worldline propagators.
In applying integration-by-parts (IBP) methods \cite{Lee:2013mka,Smirnov:2019qkx,Maierhofer:2017gsa,Klappert:2020nbg}
the presence of retarded propagators marks a 
difference, as the number of symmetries (shifts \& sign flips in loop momenta) is reduced compared to the Feynman propagator
case, which leads to larger master bases accordingly. 

\sec{Differential equations} The three-loop master integrals are solved with the method of canonical differential equations (DEs) \cite{Henn:2013pwa}. 
Grouping each master integral family 
into  a vector $\vec{I}$ one has the DE
$ 
\d\vec{I}/\d x = M(\eps, x)\, \vec{I}
$,
where $x=\gamma - \sqrt{\gamma^{2}-1}$.
We determined a transformation matrix $T(\eps,x)$
that brings us to a canonical basis 
$\vec{I}= T(\eps,x) \vec{I'}$
with an $\eps$-factorized DE 
\be
\label{eq:canA}
\d\vec{I'} = \eps A(x)\, \vec{I'} \,\d x
\ee
using the packages 
\cite{Dlapa:2022wdu,Meyer:2017joq,Prausa:2017ltv,Lee:2020zfb}.
Importantly, elliptic functions of
the first and second kind need to be included in the transformation matrix 
$T(\eps,x)$ for the J-$b$ family in order to
reach \eqref{eq:canA} \cite{Dlapa:2022wdu,Dlapa:2023hsl}.
The resulting canonical matrices $A(x)$ contain poles in $\{x,1+x,1-x,1+x^{2}\}$
--- the 4PM symbol alphabet. Expanding $\vec{I'}(\eps,x)$ in $\eps$ allows for
an iterative integration of \eqn{eq:canA} yielding logarithms and dilogarithms.

\sec{Boundary conditions} The final step is to determine the integration constants
of the $\eps$-expanded master integrals $\vec{J}$. These are fixed by taking
the static limit ($v\to 0$ i.e.~$\gamma\to 1$). We use the \emph{method of regions} \cite{Smirnov:2012gma,Bern:2021yeh,Herrmann:2021tct,Becher:2014oda} 
to expand the integrand in $v$. 
We find potential (P) and radiative (R) graviton  loop momenta
scalings as
\be
	  \ell^\text{P}_i=(\ell_i^0,\Bell_i)\sim(v,1)\,,  \qquad
	  \ell_i^\text{R}=(\ell_i^0,\Bell_i)\sim(v,v)\, .
\ee
At 4PM order there are at most two radiative gravitons:
(PP), (RR) and (PR).
This may be interpreted graphically via the example of Fig.~\ref{fig:GraphsEmergence}.
The region (PP) is  purely conservative, (PR) is purely dissipative,
and the (RR) region carries both kinds of effects.

In the (PP) region all boundary integrals reduce to test-body integrals
\cite{Jakobsen:2023ndj};
in the (RR) region they reduce to so-called double-mushroom integrals, the (PR) regions reduce to single-mushroom integrals, see Fig.~\ref{fig:GraphsEmergence}. All appearing mushroom integrals can be written as part of the $J$ family~\eqref{eq:jFamily}. This allows us to IBP reduce also the boundary integrals, which results in 10 different integral topologies; see \Fig{fig:rrGraphs}. Taking advantage of the fact that only certain linear combinations of the PP integrals are needed, we only need to calculate 17 boundary integrals explicitly to fix all integrals of all families in all regions.

Central to the evaluation of boundary integrals is the concept of \emph{active propagators},
which are sensitive to the retarded $i0^{+}$ prescription.
Graviton propagators
$1/[(\ell_{i}^{0}+i0^{+})^{2}- \Bell_{i}^{2}]$ only become active when the loop momentum
is radiative, i.e.~$\ell_{i}=\ell_i^\text{R}$, as otherwise the spatial component $\Bell_i$ dominates as $v\to 0$.
Worldline propagators, on the other hand, cannot go on-shell when directly coupled to a radiative graviton,
for otherwise the corresponding graviton integral becomes scale-free.
This property manifests itself in the IBP reduction of the boundary integrals,
wherein non-active propagators do not appear in the corresponding master integrals ---
and so, their $i0^+$ prescriptions become irrelevant.

\sec{Results}
We now present the dissipative contributions to the complete 4PM scattering observables:
\begin{align}
  \Delta X_{\rm diss}^{(4)}
  :=
  \Delta X^{(4)}-\Delta X_{\rm cons}^{(4)}
  =
  \label{eq:rad12}
  \Delta X^{(4)}_{\rm rad^1}
  +
  \Delta X^{(4)}_{\rm rad^2}\,.
\end{align}
Here $\Delta X^{(4)}$ is the full 4PM observable ---
impulse or spin kick ---
from which we subtract its conservative part $\Delta X_{\rm cons}^{(4)}$
that was already presented in Ref.~\cite{Jakobsen:2023ndj}.
We further separate dissipative terms into those
linear (rad$^1$) and quadratic (rad$^2$) in radiation,
which gain contributions only from the (PR) and (RR) regions respectively.
Dissipative observables computed within these two regions have a uniform (and opposite) behavior under the operation $v\to-v$,
under which $\gamma$ is unchanged but $\sqrt{\gamma^2-1}=\gamma v$ flips sign.

We present our observables in terms of the following functions,
all of which are strictly even under $v\to-v$:
\allowdisplaybreaks
\begin{align}\label{eq:functionBases}
  &F_{1,\ldots, 5}
  \!=\!
  \bigg\{1,
  \dfrac{\log[x]}{\sqrt{\gamma^2-1}},
  \log\! \left[\frac{\gamma_{+}}{2}\right],
  \log^2[x],
  \dfrac{\log[x]\log\!\left[\frac{\gamma_+}{2}\right]}{\sqrt{\gamma^2-1}}
  \bigg\},
  \nn
  \\
  &F_{6,\ldots,9}=
  \bigg\{
  \log[\gamma],
  \log^2\! \left[\frac{\gamma_{+}}{2}\right],
  \text{Li}_{2}\left [\frac{\gamma_{-}}{\gamma_{+}}\right],
  \text{Li}_{2}\left [-\frac{\gamma_{-}}{\gamma_{+}}\right]
  \bigg\}\,,
  \nn
  \\
  &F_{10,\ldots,13}
  =
  \bigg\{
  \dfrac{\log[x]}{\sqrt{\gamma^2-1}}\log[\gamma],
  \dfrac{\text{Li}_{2}\left [\frac{\gamma_{-}}{\gamma_{+}}\right]-4\text{Li}_{2}\left [\sqrt{\frac{\gamma_{-}}{\gamma_{+}}}\right]}{4\sqrt{\gamma^2-1}},
  \nn
  \\
  &\qquad\qquad
  \text{Li}_2[-x^{2}] -4\text{Li}_2[-x] +\log[4] \log[x] 
  -\frac{\pi^{2}}{4}
  ,
  \nn
  \\
  &\qquad\qquad
  \dfrac{\text{Li}_2[-x]-\text{Li}_2\left[-\frac{1}{x}\right]-\log[4] \log[x]}
    {\sqrt{\gamma^2-1}}
    \bigg\},
\end{align}
where $\gamma_{\pm}=\gamma\pm 1$ and $x=\gamma - \sqrt{\gamma^{2}-1}$.
The first five functions are common to both rad$^1$ and rad$^2$,
while the second line is exclusive to rad$^2$; the functions 10--13 are exclusive to rad$^1$.
In terms of these functions, the dissipative impulse and spin kick take the form
\begin{subequations}
\begin{align}
  &\Delta p_{1,\rm{diss}}^{(4)\mu} =
  \frac{m_{1}^2m_{2}^2}{|b|^{4}}
  \sum_{\alpha=1}^{13}\sum_{l=1}^{13}
  \rho_l^\mu
    F_{\alpha}
   \bigg (m_2
    d_{\alpha, l}(\gamma)
    + m_{1}\bar{d}_{\alpha, l}(\gamma)\bigg )
    ,
    \label{eq:impulsefinal}
    \\
  &\Delta S_{1,\rm{diss}}^{(4)\mu} =
  \frac{m_{1}^2m_{2}^2}{|b|^{3}}
  \sum_{\alpha=1}^{13}\sum_{l=1}^{6}
  \tilde \rho_l^\mu
    F_{\alpha}\bigg (
    m_2
    f_{\alpha,l}(\gamma)+m_1
    \bar{f}_{\alpha,l}(\gamma)\bigg)
    .
    \label{eq:spinkickfinal}
\end{align}
\end{subequations}
The basis vectors of these observables are given by:
\begin{align}\label{eq:bvtype}
     \rho^{\mu}_{l}
     &=
     \bigg\{
     \hat b^{\mu} ,
     v_{j}^{\mu} ,
     \frac{a_{i}\cdot \hat L}{|b|}\, \hat b^{\mu} ,
     \frac{a_{i}\cdot \hat L}{|b|}\,  v_{j}^{\mu} ,
     \frac{a_{i}\cdot \hat b}{|b|}\, \hat L^{\mu} ,
     \frac{a_{i}\cdot { v}_{\bar\imath}}{|b|}\, \hat L^{\mu}
     \bigg \} \,,\nn\\
     \tilde{\rho}^{\mu}_{l}
     &=
     \bigg\{
     \frac{a_1\cdot\hat{b}}{|b|}\hat{b}^\mu ,
     \frac{a_1\cdot v_2}{|b|}\hat b^{\mu} ,
     \frac{a_1\cdot\hat{b}}{|b|}v_i^\mu,
     \frac{a_1 \cdot v_2}{|b|} v_i^\mu
     \bigg\}\,,
 \end{align}
The functions $d,f$ (provided in our ancillary file) are rational functions of $\gamma$ up to integer powers of $\sqrt{\gamma^{2}-1}$. Crucially, our spin-free parts of \eqn{eq:impulsefinal} agree with \Rcites{Dlapa:2022lmu,Damgaard:2023ttc}.

We note that unlike the conservative terms~\cite{Jakobsen:2023ndj}, the dissipative ones do \emph{not} contain elliptic E/K functions or tails $\log[\frac{\gamma_-}2]$ in their functional bases~\eqref{eq:functionBases}.
Indeed the presence of tails in the conservative terms requires a cancellation of $1/\eps$ poles between the regions (PP) and (RR).
As the dissipative (PR) and (RR) regions are distinguished by their behaviour under $v\to-v$ such a cancellation cannot occur here.
The rad$^1$ parts of the dissipative observables do, however, contain 
the new functions $F_{10,\ldots,13}$ that are absent in the conservative case.
As in the conservative case, we have checked that the observables satisfy the conservation laws $p^2_i$, $S_i^2$ and $p_i\cdot S_i=0$.

\sec{Fluxes and Linear Response}
Given the complete set of scattering observables $\Delta p_1^\mu$ and $\Delta S_1^\mu$,
we can now determine the radiated linear and angular momenta,
$P_{\rm rad}^\mu$ and $J_{\rm rad}^\mu$,
over the course of the scattering --- the fluxes.
We obtain both at sub-leading order in the PM expansion:
$P_{\rm rad}^\mu$ at $\cO(G^4)$ and $J_{\rm rad}^\mu$ at $\cO(G^3)$.
Using momentum conservation the former is given trivially by
\begin{align}
  P_{\rm rad}^{(4)\mu}=-\Delta p_{1,\rm diss}^{(4)\mu}-\Delta p_{2,\rm diss}^{(4)\mu}\,,
\end{align}
with the impulse $\Delta p_2^{(4)\mu}$ obtained from $\Delta p_1^{(4)\mu}$ by simple relabelling.
Naturally, the conservative part of the impulse
$\Delta p_{1,\rm cons}^{(4)\mu}=-\Delta p_{2,\rm cons}^{(4)\mu}$ cancels,
and so $P_{\rm rad}^\mu$ gains contributions from only the dissipative parts of the impulse ---
rad$^1$ and rad$^2$.
The 4PM radiated energy $E_{\rm rad}^{(4)}=\hat{P}\cdot P_{\rm rad}^{(4)}$
agrees with the PN-expanded result~\cite{Cho:2021arx}.

To obtain $J_{\rm rad}^\mu$ at $\cO(G^3)$ we use linear response
\cite{Bini:2012ji,Bini:2021gat},
building on the work of two of the present authors~\cite{Jakobsen:2022zsx}.
Up to and including terms \emph{linear in radiation} (rad$^1$)
the dissipative parts of the scattering observables are given by the
difference between the full observables evaluated with the in-in and out-out prescriptions,
i.e.~with retarded and advanced propagators:
\begin{subequations}~\label{eq:respRels}
  \begin{align}
    &\Delta p^\mu_{i,{\rm rad}^1}=
    \sfrac12\big(\Delta p_i^\mu(J_-^\mu,p_{i-}^\mu,S_{i-}^\mu)\label{eq:consRelP}\nn
    -\Delta p_i^\mu(J_+^\mu,-p_{i+}^\mu,S_{i+}^\mu)\big)\\
    &\qquad\qquad\,\,-\cO({\rm rad}^2),\\
    &\Delta S^\mu_{i,{\rm rad}^1}=
    \sfrac12\big(\Delta S_i^\mu(J_-^\mu,p_{i-}^\mu,S_{i-}^\mu)\label{eq:consRelA}\nn
    +\Delta S_i^\mu(J_+^\mu,-p_{i+}^\mu,S_{i+}^\mu)\big)\\
    &\qquad\qquad\,\,-\cO({\rm rad}^2).
  \end{align}
\end{subequations}
At the present 4PM order, $-\cO({\rm rad}^2)$ instructs us not
to include the rad$^2$ observables $\Delta X_{{\rm rad}^2}$ on the right-hand side.
While the in-in observables are given in terms of the usual background variables
evaluated at past infinity, the out-out observables are instead given
in terms of background variables $J_+^\mu$, $p_{i+}^\mu$ and $S_{i+}^\mu$ evaluated at future infinity,
in the context of a scattering scenario where radiation is \emph{absorbed} rather than emitted.
They may be re-expressed in terms of the usual background observables using:
 ($X^{\mu}\in\{ J^{\mu}, p_{i}^{\mu}, S_{i}^{\mu}\}$)
   \begin{align}
     X_{-}^{\mu}&=X^{\mu}\, , &    X_+^{\mu}&=X^{\mu}+\Delta X_{\rm cons}^{\mu}-\Delta X_{\rm rad^1}^{\mu}\,.
   \end{align}
We have flipped the sign on the radiative components in order to reverse the direction of incoming radiation.
For later convenience, all dependence on the impact parameter $b^\mu$
has been expressed in terms of the total angular momentum vector in the center-of-mass frame $J^\mu$~\eqref{eq:jVec}.

Given our now-complete knowledge of the conservative and dissipative 4PM scattering observables,
plus all scattering observables at lower-PM orders,
the upshot is a pair of consistency requirements
that may be used to infer $J_{\rm rad}^{(3)\mu}=-\Delta J^{(3)\mu}$.
Both \eqref{eq:consRelP} and \eqref{eq:consRelA} yield the same physical constraints.
The relationships are perturbatively expandeded in $G$ to yield information at each PM order.
In practice, we make an ansatz for $\Delta J^{(3)\mu}$ on a four-dimensional basis of $\{b^\mu,L^\mu,v_1^\mu,v_2^\mu\}$,
and the linear response relations yield the $b^\mu$ and $L^\mu$ components.
The two remaining $v_i^\mu$ components are fixed by demanding that $(p_i+\Delta p_i)\cdot(L+\Delta L)=0$ up to 4PM order.

Our result for the 3PM radiated angular momentum takes the schematic form (up to
linear order in spin)
\begin{align}\label{eq:jRad}
& \!\!\! J_{\rm rad}^{(3)\mu}=\frac1E\bigg[\!-\!E^{(3)}_{\rm rad}J^\mu\!+\!\frac{\pi m_1^2m_2^2}{|b|^2}
  \sum_{i=1}^{2}\sum_{\alpha=1}^{3}\sum_{l=1}^{12}
  \rho_l^{\prime\mu}
    F_{\alpha}
    m_i
    g_{i,\alpha, l}\nn\\
    &\,\,-\frac{4m_1^2m_2^2b^\mu}{(\gamma^2-1)|b|^4}\bigg((4\gamma(2\gamma^2-1)m_1+(4\gamma^2-1)m_2)a_1\cdot v_2\nn\\
    &\,\,-(4\gamma(2\gamma^2-1)m_2+(4\gamma^2-1)m_1)a_2\cdot v_1\bigg)I(v)\bigg]\,,
\end{align}
where $E_{\rm rad}^{(3)}=\hat{P}\cdot P_{\rm rad}^{(3)}$ is the 3PM radiated energy,
and $g$ is another set of rational functions of $\gamma$ up to integer powers of $\sqrt{\gamma^{2}-1}$.
Note that we include recoil effects in all dissipative losses, so that the final value of $J^\mu$~\eqref{eq:jVec}
is defined in the final CoM frame which starting at 3PM is different from the initial CoM frame.
The function $I(v)=-\frac83+\frac1{v^2}+\frac{3v^2-1}{v^3}{\rm arccosh}\gamma$ is familiar from the 2PM
radiated angular momentum; our result is expanded on the basis
\begin{align}\label{eq:raa}
  \rho^{\prime\mu}_{l}\!=\! 
  \bigg\{\!
  \hat L^{\mu},\hat b^\mu,v_i^\mu,
  \frac{a_i\cdot\hat{L}}{|b|}\hat{L}^\mu,
  \frac{a_i\cdot\hat{b}}{|b|}\hat{b}^\mu,
  \frac{a_i\cdot v_{\bar\imath}}{|b|}v_j^\mu\bigg\}.
\end{align}
The non-spinning part of this result agrees with \Rcite{Manohar:2022dea};
for spins aligned with the orbital angular momentum vector $L^\mu$ we
agree with Refs.~\cite{Bini:2023mdz,Cho:2021arx,Cho:2021mqw} in the slow-velocity limit.
The full result~\eqref{eq:jRad} agrees with Heissenberg~\cite{carloUnpublished},
based on a very different approach using the eikonal operator, who is publishing simultaneously.

\sec{Scattering Angle} 
We consider now the total scattering angle of the relative momentum $\vct{p}$ defined in the (CoM) frame of $P^\mu$ by $p_1^\mu=(E_1,\vct{p})$ and $p_2^\mu=(E_2,-\vct{p})$.
Taking into account recoil effects its kick to this order is given as $\Delta \vct{p} = \Delta \vct{p}_1+E_1 \vct{P}_{\rm rad}/E$ and the total scattering angle (with generic spins) is given by (with $|\vct{p}|=\mu \gamma v/\Gamma$):
\begin{align}
  \cos\theta
  =
  \frac{
    \vct{p}\cdot (\vct{p}+\Delta \vct{p})
  }{
    |\vct{p}| |\vct{p}+\Delta \vct{p}|
  }
  \,.
\end{align}
We expand the angle in $G$ and up to linear order in spins:
\begin{align}
  \frac{\theta}{\Gamma}
  =
  \sum_n
  \bigg(
  \frac{G M}{|b|}
  \bigg)^n
  \Big[
    \theta^{(n,0)}
    +
    s_+
    \theta^{(n,+)}
    +
    \delta s_-
    \theta^{(n,-)}
    \Big],
\end{align}
with $s_\pm = -(a_\pm)\cdot \hat L$.
At 4PM order the angle then takes the schematic form:
\begin{align}\label{eq:angle}
  \theta^{(4,\sigma)}
  =
  \theta_{\rm{cons},\nu^0}^{(4,\sigma)}
  \!+\!
  \nu \theta_{\rm{cons},\nu}^{(4,\sigma)}
  \!+\!
  \frac{\nu}{\Gamma^2}
  \Big(
  \theta_{\rm{diss},\nu}^{(4,\sigma)}
  \!+\!
  \nu
  \theta_{\rm{diss},\nu^2}^{(4,\sigma)}
  \Big),
\end{align}
with $\sigma$ being $0$ or $\pm$.
The coefficients $\theta_{\rm{cons/diss},\nu^n}^{(4,\sigma)}$ depend only on $\gamma$ and the dissipative ones can be expanded in terms of $F_{1,\ldots,9}(\gamma)$ with polynomial coefficients up to integer powers of $\sqrt{\gamma^2-1}$.
In the spinless case our angle agrees with~\cite{Dlapa:2022lmu} and the conservative spinning terms with Ref.~\cite{Jakobsen:2023ndj}.

The first dissipative term $\theta_{\rm{diss},\nu}^{(4,\sigma)}$ gets contributions only from the (PR) region, while the second $\theta_{\rm{diss},\nu^2}^{(4,\sigma)}$ from both the (PR) and (RR) regions.
The spinless (spinning) contributions from (PR) are strictly odd (even) under the $v\to-v$ symmetry,
while the opposite is true for the dissipative (RR) contributions.
Using this symmetry the linear-in-$\nu$ terms $\theta_{\rm{cons},\nu}^{(4,\sigma)}$ and $\theta_{\rm{diss},\nu}^{(4,\sigma)}$ are uniquely defined from the full scattering angle.

From Eq.~\eqref{eq:consRelP} one may derive a linear response relation for the scattering angle:
\begin{align}
  \theta_{\rm rad^1}
  =
  -
  \frac12\left(
  \frac{\partial\theta}{\partial |J|}
  |J|_{\rm rad}+
  \frac{\partial\theta}{\partial E}
  E_{\rm rad}\right)
  +
  \mO(G^5)
  \ .
\end{align}
where the 1PM and 2PM angles and the 2PM and 3PM kicks of $|J|$ and the 3PM kick of $E$ contribute to the right-hand-side and the left-hand-side has contributions at 3PM and 4PM.
We have checked that this relation is satisfied using $J_{\rm rad}^{(3)\mu}$ from Eq.~\eqref{eq:jRad}.

Finally, we note that in the equal-mass case ($\delta=0$) the dependence of $\theta$, $J^{\mu}_{\text{rad}}$
and  $P_{\rm rad}^{\mu}$ on the spins is only through their sums, $a_1+a_2$
-- a property recently numerically observed~\cite{Rettegno:2023ghr} as well
(yet known to break at higher spin orders already at lower PM \cite{Jakobsen:2022fcj,Jakobsen:2022zsx}).

\sec{Outlook}
Having now provided a complete set of scattering observables at 4PM order including spin-orbit effects,
the obvious next step is upgrading these to include spin-spin effects.
To this end, a clear roadmap has been outlined in Refs.~\cite{Jakobsen:2022fcj,Jakobsen:2021lvp}
using the $\cN=2$ supersymmetric WQFT formalism.
Alongside this the push to 5PM order in the non-spinning case will also be vital,
initial progress having already been made in the simpler case of electrodynamics~\cite{Bern:2023ccb}.

Besides the drive towards results at higher perturbative orders in PM and spin,
also important will be resumming into the strong-field regime and mapping to bound orbits ---
with the ultimate intention of informing future waveform models.
As spin-orbit effects have already been incorporated into a resummation of the aligned-spin scattering
angle up to 3PM order~\cite{Rettegno:2023ghr},
it will be interesting to see what impact these new 4PM contributions have.
While the presence of tails continues to pose challenges for mapping 4PM results to bound orbits,
due to the presence of nonlocal-in-time effects in the conservative dynamics,
the absence of tails in the dissipative parts of our results --- and in particular the fluxes ---
leaves open the possibility of direct mappings to bound orbits.

\sec{Acknowledgments}
We thank Alessandra Buonanno, Christoph Dlapa, Kays Haddad, Johannes Henn, Gregor K\"alin, Jung-Wook Kim, Zhengwen Liu, Donal O'Connell,
Raj Patil, Rafael Porto, Chia-Hsien Shen, Jan Steinhoff 
and Paolo di Vecchia for discussions and Peter Uwer for help with high-performance computing.
We are also especially grateful to Carlo Heissenberg, who is publishing concurrently~\cite{carloUnpublished},
for a cross-check with his independent calculation of the 3PM radiated angular momentum.
This work is funded by the Deutsche Forschungsgemeinschaft
(DFG, German Research Foundation)
Projektnummer 417533893/GRK2575 ``Rethinking Quantum Field Theory''.

\newpage
\allowdisplaybreaks

\appendix
\section*{Supplementary Material}
\section{Additional integral families}
The $I^{[1]}$ and $I^{[2]}$ integral families augmenting the $J$-family~\eqref{eq:jFamily} take the form
\begin{subequations}
  \label{eq:iFamily}
  \begin{align}
    I^{[i](\sigma_1,\ldots,\sigma_6)}_{n_1,n_2,...,n_{12}}
    =
    \int_{\ell_1,\ell_2,\ell_3}\!\!\!\!\!\!
    \frac{
      \dd(\ell_1\cdot v_{i})
      \dd(\ell_2\cdot v_{1})
      \dd(\ell_3\cdot v_{1})
    }{
      D_1^{n_1}
      D_2^{n_2}
      ...
      D_{12}^{n_{12}}
    },
  \end{align}
  with the propagators ($j=1,2,3$ and $k=1,2$):
  \begin{align}
    &D_{1}
    =
    \ell_1\cdot v_{\bar\imath} +\sigma_1i0^+
    \,,\,\,
    D_{1+k}
    =
    \ell_{1+k}\cdot v_2 +\sigma_{1+k}i0^+\,,\nn
    \\
    &D_{4}
    =
    (\ell_1+\ell_2+\ell_3+q)^2+\sigma_{4}\,\text{sgn}(\ell_1^0+\ell_2^0+\ell_3^0)\,i0^+
    \,,\,\, \nn\\
    &D_{5}
    =
    (\ell_1+\ell_2+q)^2+\sigma_{5}\,\text{sgn}(\ell_1^0+\ell_2^0)\,i0^+
    \,,
    \\
    &D_{6}\!
    =
    (\ell_1+\ell_3)^2+\sigma_{6}\,\text{sgn}(\ell_1^0+\ell_3^0)\,i0^+
    \,,\,\,\nn\\
    &D_{7}\!
    =
    (\ell_2+\ell_3)^2
    \,,\,\,
    D_{7+j}
    =
    \ell_{j}^2
    \,,\,\,
    D_{10+k}\!
    =
    (\ell_{k}+q)^2\, ,\nn
  \end{align}
\end{subequations}
In the $I^{[1]}$ integral family, corresponding to the test-body integrals,
all graviton propagators may be considered passive, and so $\sigma_4$--$\sigma_6$ may be safely ignored.
In the comparable-mass $I^{[2]}$ family propagators $D_5$ and $D_6$ do not appear in any denominators that occur,
and so $\sigma_5$ and $\sigma_6$ are unimportant.

\bibliographystyle{JHEP}
\bibliography{spin-4PM-dissipative}

\end{document}

%% file: Graphs_Emergence_RRPP.tex
\resizebox{0.8\textwidth}{!}{


\begin{tikzpicture}[baseline={([yshift=3ex]current bounding box.south)},scale=.7]
  \coordinate (inA) at (0.4,.7);
  \coordinate (outA) at (5.35,.7);
  \coordinate (inB) at (0.4,-.7);
  \coordinate (outB) at (5.35,-.7);
  \coordinate (xA) at (1,.7);
  \coordinate (xxA) at (1.55,.7) ;
  \coordinate (lA) at (2.25,.7) ;
  \coordinate (rA) at (3.5,.7) ;
   \coordinate (lB) at (2.25,-.7) ;
  \coordinate (rB) at (3.5,-.7) ;
  \coordinate (yA) at (2.60,.7);
  \coordinate (yyA) at (3.15,.7) ;
  \coordinate (zA) at (4.20,.7);
  \coordinate (zzA) at (4.75,.7) ;
  \coordinate (xB) at (1,-.7);
   \coordinate (xxB) at (1.55,-.7) ;
  \coordinate (yB) at (2.5,-.7);
   \coordinate (yyB) at (3.15,-.7) ;
  \coordinate (zB) at (4,-.7);
  \coordinate (zzB) at (4.75,-.7);
  \coordinate (xM) at (1,0);
  \coordinate (yM) at (2.5,0);
  \coordinate (zM) at (4,-0);
  \draw [dotted] (inA) -- (outA);
  \draw [dotted] (inB) -- (outB);
  \draw [draw=none] (xA) to[out=40,in=140] (zA);
  \draw [photon] (xA) -- (xB);
  \draw [photon] (zzA) -- (zzB);
  \draw [photon] (lA) -- (lB);
  \draw [photon] (rA) --  (rB);
  \draw [zUndirected] (xA) -- (lA);
  \draw [zUndirected] (lB) -- (rB);
  \draw [zUndirected] (rA) -- (zzA);
  \draw [fill] (xA) circle (.08);
  \draw [fill] (lA) circle (.08);
  \draw [fill] (rA) circle (.08);
  \draw [fill] (zzA) circle (.08);
 \draw [fill] (xB) circle (.08);
  \draw [fill] (lB) circle (.08);
  \draw [fill] (rB) circle (.08);
  \draw [fill] (zzB) circle (.08);
\end{tikzpicture}

$\longrightarrow$

\begin{tikzpicture}[baseline={([yshift=3ex]current bounding box.south)},scale=.7]
  \coordinate (inA) at (0.4,.7);
  \coordinate (outA) at (5.35,.7);
  \coordinate (inB) at (0.4,-.7);
  \coordinate (outB) at (5.35,-.7);
  \coordinate (xA) at (1,.7);
  \coordinate (xxA) at (1.55,.7) ;
  \coordinate (lA) at (2.25,.7) ;
  \coordinate (rA) at (3.5,.7) ;
   \coordinate (lB) at (2.25,-.7) ;
  \coordinate (rB) at (3.5,-.7) ;
  \coordinate (yA) at (2.60,.7);
  \coordinate (yyA) at (3.15,.7) ;
  \coordinate (zA) at (4.20,.7);
  \coordinate (zzA) at (4.75,.7) ;
  \coordinate (xB) at (1,-.7);
   \coordinate (xxB) at (1.55,-.7) ;
  \coordinate (yB) at (2.5,-.7);
   \coordinate (yyB) at (3.15,-.7) ;
  \coordinate (zB) at (4,-.7);
  \coordinate (zzB) at (4.75,-.7);
  \coordinate (xM) at (1,0);
  \coordinate (yM) at (2.5,0);
  \coordinate (zM) at (4,-0);
  \draw [dotted] (inA) -- (outA);
  \draw [dotted] (inB) -- (outB);
  \draw [draw=none] (xA) to[out=40,in=140] (zA);
  \draw [photon] (xA) -- (xB);
  \draw [photon] (zzA) -- (zzB);
   \draw [photon] (lA) -- (lB);
  \draw [photon] (rA) --  (rB);
  \draw [zUndirected, color=red] (xA) -- (lA);
  \draw [zUndirected, color=red] (lA) -- (rA);
  \draw [zUndirected, color=red] (rA) -- (zzA);
  \draw [fill] (xA) circle (.08);
  \draw [fill] (lA) circle (.08);
  \draw [fill] (rA) circle (.08);
  \draw [fill] (zzA) circle (.08);
 \draw [fill] (xB) circle (.08);
  \draw [fill] (lB) circle (.08);
  \draw [fill] (rB) circle (.08);
  \draw [fill] (zzB) circle (.08);
   \draw (2.9,.9) node [above] {PP};
\end{tikzpicture}

$+$

\begin{tikzpicture}[baseline={([yshift=3ex]current bounding box.south)},scale=.7]
  \coordinate (inA) at (0.4,.7);
  \coordinate (outA) at (5.35,.7);
  \coordinate (inB) at (0.4,-.7);
  \coordinate (outB) at (5.35,-.7);
  \coordinate (xA) at (1,.7);
  \coordinate (xxA) at (1.55,.7) ;
  \coordinate (yA) at (2.60,.7);
  \coordinate (yyA) at (3.15,.7) ;
  \coordinate (zA) at (4.20,.7);
  \coordinate (zzA) at (4.75,.7) ;
  \coordinate (xB) at (1,-.7);
  \coordinate (yB) at (2.5,-.7);
  \coordinate (zB) at (4,-.7);
  \coordinate (zzB) at (4.75,-.7);
  \coordinate (xM) at (1,0);
  \coordinate (yM) at (2.5,0);
  \coordinate (zM) at (4,-0);
  \draw [dotted] (inA) -- (outA);
  \draw [dotted] (inB) -- (outB);
  \draw [draw=none] (xA) to[out=40,in=140] (zA);
  \draw [photon] (xA) -- (xB);
  \draw [photon] (zzA) -- (zzB);
  \draw [photon, color=red] (xxA) to[out=-90,in=-90] (yA);
  \draw [photon, color=red] (yyA) to[out=-80,in=-100] (zA);
  \draw [zUndirected] (xA) -- (xxA);
  \draw [zUndirected] (yA) -- (yyA);
  \draw [zUndirected] (zA) -- (zzA);
  \draw [fill] (xA) circle (.08);
  \draw [fill] (xxA) circle (.08);
  \draw [fill] (yA) circle (.08);
  \draw [fill] (yyA) circle (.08);
  \draw [fill] (zA) circle (.08);
  \draw [fill] (zzA) circle (.08);
  \draw [fill] (xB) circle (.08);
  \draw [fill] (zzB) circle (.08);
   \draw (2.9,.9) node [above] {RR};
\end{tikzpicture}

$+$

\begin{tikzpicture}[baseline={([yshift=3ex]current bounding box.south)},scale=.7]
  \coordinate (inA) at (0.4,.7);
  \coordinate (outA) at (5.35,.7);
  \coordinate (inB) at (0.4,-.7);
  \coordinate (outB) at (5.35,-.7);
  \coordinate (xA) at (1,.7);
  \coordinate (xxA) at (1.55,.7) ;
  \coordinate (yA) at (2.50,.7);
  \coordinate (yyA) at (3.15,.7) ;
  \coordinate (zA) at (4.20,.7);
  \coordinate (zzA) at (4.75,.7) ;
  \coordinate (xB) at (1,-.7);
  \coordinate (yB) at (2.5,-.7);
  \coordinate (zB) at (4,-.7);
  \coordinate (zzB) at (4.75,-.7);
  \coordinate (xM) at (1,0);
  \coordinate (yM) at (2.5,0);
  \coordinate (zM) at (4,-0);
  \draw [dotted] (inA) -- (outA);
  \draw [dotted] (inB) -- (outB);
  \draw [draw=none] (xA) to[out=40,in=140] (zA);
  \draw [photon] (xA) -- (xB);
  \draw [photon] (zzA) -- (zzB);
  \draw [photon] (yB) -- (yA);
  \draw [photon, color=red] (yyA) to[out=-80,in=-100] (zA);
  \draw [zUndirected, color=red] (xA) -- (yA);
  \draw [zUndirected] (yA) -- (yyA);
  \draw [zUndirected] (zA) -- (zzA);
  \draw [fill] (xA) circle (.08);
  \draw [fill] (yB) circle (.08);
  \draw [fill] (yA) circle (.08);
  \draw [fill] (yyA) circle (.08);
  \draw [fill] (zA) circle (.08);
  \draw [fill] (zzA) circle (.08);
  \draw [fill] (xB) circle (.08);
  \draw [fill] (zzB) circle (.08);
   \draw (2.9,.9) node [above] {PR};
\end{tikzpicture}

$+$

\begin{tikzpicture}[baseline={([yshift=3ex]current bounding box.south)},scale=.7]
  \coordinate (inA) at (0.4,.7);
  \coordinate (outA) at (5.35,.7);
  \coordinate (inB) at (0.4,-.7);
  \coordinate (outB) at (5.35,-.7);
  \coordinate (xA) at (1,.7);
  \coordinate (xxA) at (1.55,.7) ;
  \coordinate (yA) at (2.60,.7);
  \coordinate (yyA) at (3.15,.7) ;
  \coordinate (zA) at (4.20,.7);
  \coordinate (zzA) at (4.75,.7) ;
  \coordinate (xB) at (1,-.7);
  \coordinate (yB) at (2.5,-.7);
  \coordinate (zB) at (4,-.7);
  \coordinate (zzB) at (4.75,-.7);
  \coordinate (xM) at (1,0);
  \coordinate (yM) at (2.5,0);
  \coordinate (zM) at (4,-0);
  \draw [dotted] (inA) -- (outA);
  \draw [dotted] (inB) -- (outB);
  \draw [draw=none] (xA) to[out=40,in=140] (zA);
  \draw [photon] (xA) -- (xB);
  \draw [photon] (zzA) -- (zzB);
  \draw [photon, color=red] (xxA) to[out=-90,in=-90] (yA);
  \draw [photon] (yyA) --  (yyB);
  \draw [zUndirected] (xA) -- (xxA);
  \draw [zUndirected] (yA) -- (yyA);
  \draw [zUndirected, color=red] (yyA) -- (zzA);
  \draw [fill] (xA) circle (.08);
  \draw [fill] (xxA) circle (.08);
  \draw [fill] (yA) circle (.08);
  \draw [fill] (yyA) circle (.08);
  \draw [fill] (yyB) circle (.08);
  \draw [fill] (zzA) circle (.08);
  \draw [fill] (xB) circle (.08);
  \draw [fill] (zzB) circle (.08);
   \draw (2.9,.9) node [above] {RP};
\end{tikzpicture}

}

%% file: RRP_Graph.tex
\resizebox{1\textwidth}{!}{\begin{tikzpicture}[baseline={([yshift=-1ex]current bounding box.south)},scale=.7]
  \coordinate (inA) at (0.4,.7);
  \coordinate (outA) at (4.35,.7);
  \coordinate (inB) at (0.4,-.7);
  \coordinate (outB) at (4.35,-.7);
  \coordinate (xA) at (1,.7);
  \coordinate (xxA) at (1.825,.7) ;
  \coordinate (xxB) at (1.825,-.7) ;
  \coordinate (yA) at (1.60,.7);
  \coordinate (yyA) at (2.375,.7) ;
  \coordinate (zA) at (2.925,.7);
  \coordinate (zzA) at (3.75,.7) ;
  \coordinate (xB) at (1,-.7);
  \coordinate (yB) at (1.5,-.7);
  \coordinate (zB) at (2.925,-.7);
  \coordinate (zzB) at (3.75,-.7);
  \coordinate (xM) at (1,0);
  \coordinate (yM) at (1.5,0);
  \coordinate (zM) at (3,-0);
  \draw [dotted] (inA) -- (outA);
  \draw [dotted] (inB) -- (outB);
  \draw [draw=none] (xA) to[out=40,in=140] (zA);
  \draw [photon] (yyA) -- (xB);
  \draw [photon] (yyA) -- (zB);
  \draw [photon] (yyA) -- (zzB);
  \draw [photon] (yyA) -- (xxB);

  \draw [fill] (xxB) circle (.08);

  \draw [fill] (yyA) circle (.08);

  \draw [fill] (zB) circle (.08);
  \draw [fill] (xB) circle (.08);
  \draw [fill] (zzB) circle (.08);

\end{tikzpicture}
\begin{tikzpicture}[baseline={([yshift=-1ex]current bounding box.south)},scale=.7]
  \coordinate (inA) at (0.4,.7);
  \coordinate (outA) at (4.35,.7);
  \coordinate (inB) at (0.4,-.7);
  \coordinate (outB) at (4.35,-.7);
  \coordinate (xA) at (1,.7);
  \coordinate (xxA) at (1.825,.7) ;
  \coordinate (xxB) at (1.825,-.7) ;
  \coordinate (yA) at (1.60,.7);
  \coordinate (yyA) at (2.375,.7) ;
  \coordinate (zA) at (2.925,.7);
  \coordinate (zzA) at (3.75,.7) ;
  \coordinate (xB) at (1,-.7);
  \coordinate (yB) at (1.5,-.7);
  \coordinate (zB) at (2.925,-.7);
  \coordinate (zzB) at (3.75,-.7);
  \coordinate (xM) at (1,0);
  \coordinate (yM) at (1.5,0);
  \coordinate (zM) at (3,-0);
  \draw [dotted] (inA) -- (outA);
  \draw [dotted] (inB) -- (outB);
  \draw [draw=none] (xA) to[out=40,in=140] (zA);
  \draw [photon] (xA) -- (xB);
  \draw [photon] (yyA) -- (zB);
  \draw [photon] (zzA) -- (zzB);
  \draw [photon] (yyA) -- (xxB);
  \draw [zUndirected,color=red] (yyA) -- (xA);
   \draw [zUndirected,color=red] (yyA) -- (zzA);
  \draw [fill] (xxB) circle (.08);

  \draw [fill] (yyA) circle (.08);
  \draw [fill] (xA) circle (.08);
  \draw [fill] (zzA) circle (.08);
  \draw [fill] (zB) circle (.08);
  \draw [fill] (xB) circle (.08);
  \draw [fill] (zzB) circle (.08);

\end{tikzpicture}
\begin{tikzpicture}[baseline={([yshift=-1ex]current bounding box.south)},scale=.7]
  \coordinate (inA) at (0.4,.7);
  \coordinate (outA) at (4.35,.7);
  \coordinate (inB) at (0.4,-.7);
  \coordinate (outB) at (4.35,-.7);
  \coordinate (xA) at (1,.7);
  \coordinate (xxA) at (1.825,.7) ;
  \coordinate (xxB) at (1.825,-.7) ;
  \coordinate (yA) at (1.60,.7);
  \coordinate (yyA) at (1.875,.7) ;
  \coordinate (yyB) at (1.875,-.7) ;
  \coordinate (zA) at (2.925,.7);
  \coordinate (zzA) at (3.3375,.7) ;
  \coordinate (xB) at (1,-.7);
  \coordinate (yB) at (1.5,-.7);
  \coordinate (zB) at (2.925,-.7);
  \coordinate (zzB) at (3.75,-.7);
  \coordinate (xM) at (1,0);
  \coordinate (yM) at (2.5,0);
  \coordinate (zM) at (3,-0);
  \draw [dotted] (inA) -- (outA);
  \draw [dotted] (inB) -- (outB);
  \draw [draw=none] (xA) to[out=40,in=140] (zA);
  \draw [photon] (xA) -- (xB);
  \draw [photon] (yyA) -- (yyB);
  \draw [photon] (zzA) -- (zzB);
  \draw [photon] (zB) -- (zzA);
  \draw [zUndirected,color=red] (yyA) -- (xA);
   \draw [zUndirected,color=red] (yyA) -- (zzA);
  \draw [fill] (yyB) circle (.08);

  \draw [fill] (yyA) circle (.08);
  \draw [fill] (xA) circle (.08);
  \draw [fill] (zzA) circle (.08);
  \draw [fill] (zB) circle (.08);
  \draw [fill] (xB) circle (.08);
  \draw [fill] (zzB) circle (.08);

\end{tikzpicture}
\begin{tikzpicture}[baseline={([yshift=-1ex]current bounding box.south)},scale=.7]
  \coordinate (inA) at (0.4,.7);
  \coordinate (outA) at (4.35,.7);
  \coordinate (inB) at (0.4,-.7);
  \coordinate (outB) at (4.35,-.7);
  \coordinate (xA) at (1,.7);
  \coordinate (xxA) at (1.825,.7) ;
  \coordinate (xxB) at (1.825,-.7) ;
  \coordinate (yA) at (1.60,.7);
  \coordinate (yyA) at (1.875,.7) ;
  \coordinate (yyB) at (2.1,-.7) ;
  \coordinate (zA) at (2.925,.7);
  \coordinate (zzA) at (3.3375,.7) ;
  \coordinate (xB) at (1,-.7);
  \coordinate (yB) at (1.5,-.7);
  \coordinate (zB) at (2.925,-.7);
  \coordinate (zzB) at (3.75,-.7);
  \coordinate (xM) at (1,0);
  \coordinate (yM) at (1.5,0);
  \coordinate (zM) at (3,-0);
  \draw [dotted] (inA) -- (outA);
  \draw [dotted] (inB) -- (outB);
  \draw [draw=none] (xA) to[out=40,in=140] (zA);
  \draw [photon] (xA) -- (xB);
  \draw [photon] (zA) -- (yyB);
  \draw [photon] (zA) -- (zzB);
  \draw [photon] (zB) -- (zA);
  \draw [zUndirected,color=red] (zA) -- (xA);

  \draw [fill] (yyB) circle (.08);

  \draw [fill] (xA) circle (.08);
  \draw [fill] (zA) circle (.08);
  \draw [fill] (zB) circle (.08);
  \draw [fill] (xB) circle (.08);
  \draw [fill] (zzB) circle (.08);

\end{tikzpicture}
\begin{tikzpicture}[baseline={([yshift=-1ex]current bounding box.south)},scale=.7]
  \coordinate (inA) at (0.4,.7);
  \coordinate (outA) at (4.35,.7);
  \coordinate (inB) at (0.4,-.7);
  \coordinate (outB) at (4.35,-.7);
  \coordinate (xA) at (1.9,.7);
  \coordinate (xxA) at (1.3375,.7) ;
  \coordinate (xxB) at (1.3375,-.7) ;
  \coordinate (yA) at (1.60,.7);
  \coordinate (yyA) at (1.875,.7) ;
  \coordinate (yyB) at (1.7375,-.7) ;
  \coordinate (zA) at (2.925,.7);
  \coordinate (zzA) at (3.3375,.7) ;
  \coordinate (xB) at (0.9375,-.7);
  \coordinate (yB) at (1.5,-.7);
  \coordinate (zB) at (2.9375,-.7);
  \coordinate (zzB) at (3.75,-.7);
  \coordinate (xM) at (1,0);
  \coordinate (yM) at (2.5,0);
  \coordinate (zM) at (4,-0);
  \draw [dotted] (inA) -- (outA);
  \draw [dotted] (inB) -- (outB);
  \draw [draw=none] (xA) to[out=40,in=140] (zA);
  \draw [photon] (xxA) -- (xB);
  \draw [photon] (xxA) -- (yyB);
  \draw [photon] (zzA) -- (zzB);
  \draw [photon] (zB) -- (zzA);
  \draw [zUndirected,color=red] (xxA) -- (zzA);

  \draw [fill] (yyB) circle (.08);

  \draw [fill] (xxA) circle (.08);
  \draw [fill] (zzA) circle (.08);
  \draw [fill] (zB) circle (.08);
  \draw [fill] (xB) circle (.08);
  \draw [fill] (zzB) circle (.08);

\end{tikzpicture}
\begin{tikzpicture}[baseline={([yshift=-1ex]current bounding box.south)},scale=.7]
  \coordinate (inA) at (0.4,.7);
  \coordinate (outA) at (4.35,.7);
  \coordinate (inB) at (0.4,-.7);
  \coordinate (outB) at (4.35,-.7);
  \coordinate (xA) at (1,.7);
  \coordinate (xxA) at (2,.7) ;
  \coordinate (xxB) at (2,-.7) ;
  \coordinate (yA) at (3,.7);
  \coordinate (yB) at (3,-.7);
  \coordinate (zA) at (4,.7);
  \coordinate (zB) at (4,-.7);
  \draw [dotted] (inA) -- (outA);
  \draw [dotted] (inB) -- (outB);
  \draw [draw=none] (xA) to[out=40,in=140] (zA);
  \draw [photon] (xA) -- (xB);
  \draw [photon] (xxA) -- (xxB);
  \draw [photon] (yA) -- (yB);
  \draw [photon] (zA) -- (zB);
  \draw [zUndirected,color=red] (xA) -- (xxA);
  \draw [zUndirected,color=red] (xxA) -- (yA);
  \draw [zUndirected,color=red] (xxA) -- (zA);

  \draw [fill] (xA) circle (.08);
  \draw [fill] (xxA) circle (.08);
  \draw [fill] (yA) circle (.08);
  \draw [fill] (zA) circle (.08);
  \draw [fill] (xB) circle (.08);
  \draw [fill] (xxB) circle (.08);
  \draw [fill] (yB) circle (.08);
  \draw [fill] (zB) circle (.08);

\end{tikzpicture}
\begin{tikzpicture}[baseline={([yshift=-1ex]current bounding box.south)},scale=.7]
  \coordinate (inA) at (0.4,.7);
  \coordinate (outA) at (4.35,.7);
  \coordinate (inB) at (0.4,-.7);
  \coordinate (outB) at (4.35,-.7);
  \coordinate (xA) at (1,.7);
  \coordinate (xxA) at (2,.7) ;
  \coordinate (xxB) at (2,-.7) ;
  \coordinate (yA) at (3,.7);
  \coordinate (yB) at (3,-.7);
  \coordinate (zA) at (4,.7);
  \coordinate (zB) at (4,-.7);
  \draw [dotted] (inA) -- (outA);
  \draw [dotted] (inB) -- (outB);
  \draw [draw=none] (xA) to[out=40,in=140] (zA);
  \draw [photon] (xA) -- (xB);
  \draw [photon] (xxA) -- (xxB);
  \draw [photon] (yA) -- (yB);
  \draw [photon] (zA) -- (zB);
  \draw [zUndirected,color=red] (xA) to[out=45,in=135](yA);
  \draw [zUndirected,color=red] (xxA) -- (yA);
  \draw [zUndirected,color=red] (xxA) -- (zA);

  \draw [fill] (xA) circle (.08);
  \draw [fill] (xxA) circle (.08);
  \draw [fill] (yA) circle (.08);
  \draw [fill] (zA) circle (.08);
  \draw [fill] (xB) circle (.08);
  \draw [fill] (xxB) circle (.08);
  \draw [fill] (yB) circle (.08);
  \draw [fill] (zB) circle (.08);

\end{tikzpicture}
\begin{tikzpicture}[baseline={([yshift=-1ex]current bounding box.south)},scale=.7]
  \coordinate (inA) at (0.4,.7);
  \coordinate (outA) at (5.35,.7);
  \coordinate (inB) at (0.4,-.7);
  \coordinate (outB) at (5.35,-.7);
  \coordinate (xA) at (1,.7);
  \coordinate (xxA) at (1.55,.7) ;
  \coordinate (yA) at (2.60,.7);
  \coordinate (yyA) at (2.875,.7) ;
  \coordinate (zA) at (4.20,.7);
  \coordinate (zzA) at (4.75,.7) ;
  \coordinate (xB) at (1,-.7);
  \coordinate (yB) at (2.5,-.7);
  \coordinate (zB) at (4,-.7);
  \coordinate (zzB) at (4.75,-.7);
  \coordinate (xM) at (1,0);
  \coordinate (yM) at (2.5,0);
  \coordinate (zM) at (4,-0);
  \draw [dotted] (inA) -- (outA);
  \draw [dotted] (inB) -- (outB);
  \draw [draw=none] (xA) to[out=40,in=140] (zA);
  \draw [photon] (xA) -- (xB);
  \draw [photon] (zzA) -- (zzB);
  \draw [photon, color=red] (xA) to[out=-90,in=-90] (yyA);
  \draw [photon, color=red] (yyA) to[out=-80,in=-100] (zzA);

  \draw [fill] (xA) circle (.08);

  \draw [fill] (yyA) circle (.08);

  \draw [fill] (zzA) circle (.08);
  \draw [fill] (xB) circle (.08);
  \draw [fill] (zzB) circle (.08);

\end{tikzpicture}

 \begin{tikzpicture}[baseline={([yshift=-1ex]current bounding box.south)},scale=.7]
  \coordinate (inA) at (0.4,.7);
  \coordinate (outA) at (4.85,.7);
  \coordinate (inB) at (0.4,-.7);
  \coordinate (outB) at (4.85,-.7);
  \coordinate (xA) at (1,.7);
  \coordinate (xxA) at (1.70,.7) ;
  \coordinate (yA) at (2.875,.7) ;
  \coordinate (zA) at (4.25,.7) ;
  \coordinate (xB) at (1,-.7);
  \coordinate (yB) at (2.875,-.7);
    \coordinate (zB) at (4.25,-.7);

    \draw [fill] (xA) circle (.08);
    \draw [fill] (yA) circle (.08);

    \draw [fill] (xB) circle (.08);
    \draw [fill] (yB) circle (.08);
    \draw [fill] (zB) circle (.08);
    \draw [dotted] (inA) -- (outA);
    \draw [dotted] (inB) -- (outB);
    \draw [draw=none] (xA) to[out=40,in=140] (zA);
    \draw [photon] (xA) -- (xB);
    \draw [photon] (yA) -- (yB);
    \draw [photon] (yA) -- (zB);
    \draw [photon, color=red] (xA) to[out=-80,in=-100] (yA);

  \end{tikzpicture}

 \begin{tikzpicture}[baseline={([yshift=-1ex]current bounding box.south)},scale=.7]
  \coordinate (inA) at (0.4,.7);
  \coordinate (outA) at (5.35,.7);
  \coordinate (inB) at (0.4,-.7);
  \coordinate (outB) at (5.35,-.7);
  \coordinate (xA) at (1,.7);
  \coordinate (xxA) at (1.70,.7) ;
  \coordinate (yA) at (2.875,.7) ;
  \coordinate (zA) at (4.75,.7) ;
  \coordinate (xB) at (1,-.7);
  \coordinate (yB) at (2.875,-.7);
    \coordinate (zB) at (4.75,-.7);

    \draw [fill] (xA) circle (.08);
    \draw [fill] (yA) circle (.08);
    \draw [fill] (zA) circle (.08);
    \draw [fill] (xB) circle (.08);
    \draw [fill] (yB) circle (.08);
    \draw [fill] (zB) circle (.08);
    \draw [dotted] (inA) -- (outA);
    \draw [dotted] (inB) -- (outB);
    \draw [draw=none] (xA) to[out=40,in=140] (zA);
    \draw [photon] (xA) -- (xB);
    \draw [photon] (yA) -- (yB);
    \draw [photon] (zA) -- (zB);
    \draw [photon, color=red] (xA) to[out=-80,in=-100] (yA);
    \draw [zUndirected,color=red] (yA) -- (zA);
  \end{tikzpicture}

}